%% file: article.tex
\documentclass[acmsmall, nonacm, screen]{acmart}

\AtBeginDocument{%
  \providecommand\BibTeX{{%
    \normalfont B\kern-0.5em{\scshape i\kern-0.25em b}\kern-0.8em\TeX}}}

\setcopyright{acmcopyright}
\copyrightyear{2018}
\acmYear{2018}
\acmDOI{10.1145/1122445.1122456}

\acmConference[Woodstock '18]{Woodstock '18: ACM Symposium on Neural
  Gaze Detection}{June 03--05, 2018}{Woodstock, NY}
\acmBooktitle{Woodstock '18: ACM Symposium on Neural Gaze Detection,
  June 03--05, 2018, Woodstock, NY}
\acmPrice{15.00}
\acmISBN{978-1-4503-XXXX-X/18/06}

\begin{document}

\title[Design of a Privacy-Preserving Data Platform for Collaboration Against Human Trafficking]{Design of a Privacy-Preserving Data Platform for Collaboration Against Human Trafficking}

\author{Darren Edge}
\email{daedge@microsoft.com}
\affiliation{%
  \institution{Microsoft Research}
  \city{Cambridge}
  \country{UK}
}

\author{Weiwei Yang}
\email{weiwya@microsoft.com}
\affiliation{%
  \institution{Microsoft Research}
  \city{Redmond}
  \state{WA}
  \country{USA}
}

\author{Kate Lytvynets}
\email{kalytv@microsoft.com}
\affiliation{%
  \institution{Microsoft Research}
  \city{Redmond}
  \state{WA}
  \country{USA}
}

\author{Harry Cook}
\authornote{The opinions expressed in the report are those of the authors and do not necessarily reflect the views of the International Organization for Migration (IOM). The designations employed and the presentation of material throughout the report do not imply expression of any opinion whatsoever on the part of IOM concerning legal status of any country, territory, city or area, or of its authorities, or concerning its frontiers or boundaries.}
\email{hcook@iom.int}
\affiliation{%
  \institution{IOM (UN Migration)}
  \city{Geneva}
  \country{Switzerland}
}

\author{Claire Galez-Davis}
\authornotemark[1]
\email{cgalez@iom.int}
\affiliation{%
  \institution{IOM (UN Migration)}
  \city{Geneva}
  \country{Switzerland}
}

\author{Hannah Darnton}
\email{hdarnton@bsr.org}
\affiliation{%
  \institution{Business for Social Responsibility (BSR)}
  \city{San Francisco}
  \state{CA}
  \country{USA}
}

\author{Christopher White}
\email{chwh@microsoft.com}
\affiliation{%
  \institution{Microsoft Research}
  \city{Redmond}
  \state{WA}
  \country{USA}
}

\renewcommand{\shortauthors}{Working paper, Edge et al. 2020}

\begin{abstract}
Case records on victims of human trafficking are highly sensitive, yet the ability to share such data is critical to evidence-based practice and policy development across government, business, and civil society. We present new methods to anonymize, publish, and explore such data, implemented as a pipeline generating three artifacts:  (1) synthetic data mitigating the privacy risk that published attribute combinations might be linked to known individuals or groups; (2) aggregate data mitigating the utility risk that synthetic data might misrepresent statistics needed for official reporting; and (3) visual analytics interfaces to both datasets mitigating the accessibility risk that privacy mechanisms or analysis tools might not be understandable and usable by all stakeholders. We present our work as a design study motivated by the goal of transforming how the world's largest database of identified victims is made available for global collaboration against human trafficking.
\end{abstract}




\maketitle

\input{introduction.tex}
\input{related_work.tex}
\input{discover.tex}
\input{design.tex}
\input{implementation.tex}
\input{discussion.tex}
\input{conclusion.tex}

\bibliographystyle{ACM-Reference-Format}
\bibliography{references.bib}

\end{document}

%% file: introduction.tex
\section{Introduction}

Human trafficking is a complex crime with a foothold in every country. While the available data are sparse and there is no global estimate of overall prevalence, ILO, IOM, and the Walk Free Foundation estimated that the related crimes of forced labor and forced marriage had as many as 40 million global victims in 2016 \cite{ILO}. Much effort has been invested in the identification and investigation of individual cases \cite{deeb2019understanding}, with notable tools including TellFinder \cite{halltellfinder}, DIG \cite{kejriwal2018technology}, and Traffic Jam \cite{trafficjam} for linking and querying online ads for commercial sex, Freedom Signal \cite{SAS} for posting decoy sex ads and deterrence chatbots, Spotlight \cite{spotlight} for supporting end-to-end juvenile sex trafficking investigations, and the Apprise mobile application \cite{10.1145/3290605.3300385} for victim identification. Recent years have also seen the field of HCI take a shift towards more socially complex topics \cite{10.1145/3025453.3025615}, including human trafficking \cite{kejriwal2018technology, 10.1145/3290605.3300385, deeb2019understanding} and extending to the related issues of surviving intimate partner abuse \cite{10.1145/3025453.3025875}, providing social justice for sex workers \cite{10.1145/3025453.3025615}, and designing for fairness, accountability, and transparency in public sector decision making \cite{10.1145/3173574.3174014}. In all of these cases, there is an urgent need to ensure that vulnerable populations are both \emph{represented} to those making evidence-based policy decisions while also being \emph{protected} from those who would cause them harm.

For the case of human trafficking in particular, the 2019 Trafficking in Persons report \cite{tip2019} describes the many challenges to building and sharing datasets that facilitate collaboration between governments and civil society. These include the need for trauma-informed data collection as well as appropriate data standardization and anonymization to protect the vulnerable individuals represented in published datasets. Despite the challenges, the report identifies the Counter Trafficking Data Collaborative (CTDC) \cite{CTDC} as a benchmark initiative in the collection, management, and dissemination of  human trafficking case data. Launched in 2017, CTDC combines victim case records from IOM (UN Migration), Polaris, and Liberty Shared to create the world's largest database of its kind and an online data platform through which derived data artifacts can be published.

In July 2019, CTDC joined the accelerator program of the technology industry coalition Tech Against Trafficking (TAT) \cite{TAT}, with the joint goals of advancing the privacy, utility, and accessibility of the CTDC data platform. Over the course of this accelerator, we have worked with CTDC and the broader community to achieve these goals, making the following high-level contributions:

\begin{enumerate}
    \item \emph{Privacy-preserving algorithm} -- we developed the concept of \emph{k-synthetic anonymity} and an algorithm for achieving it. This concept generalizes the  notion of $k$-anonymity \cite{sweeney2002k, sweeney2002achieving} to all columns of a sensitive dataset, requiring the generation of synthetic data records whose attribute combinations always describe groups of at least $k$ individuals in the sensitive dataset.
    \item \emph{Privacy-preserving interface} -- we designed a new approach to privacy-preserving visual analytics in which ``estimated'' counts observed while exploring synthetic data are juxtaposed with precomputed ``actual'' counts that have been rounded and thresholded for release.
    \item \emph{Privacy-preserving pipeline} -- we implemented and released an open-source Python pipeline for generating synthetic data, aggregate data, and visual analytics interfaces from any sensitive microdata (i.e., data in which each record represents an individual). 
\end{enumerate}

\begin{figure}[ph]
  \centering
  \includegraphics[width=13.5cm]{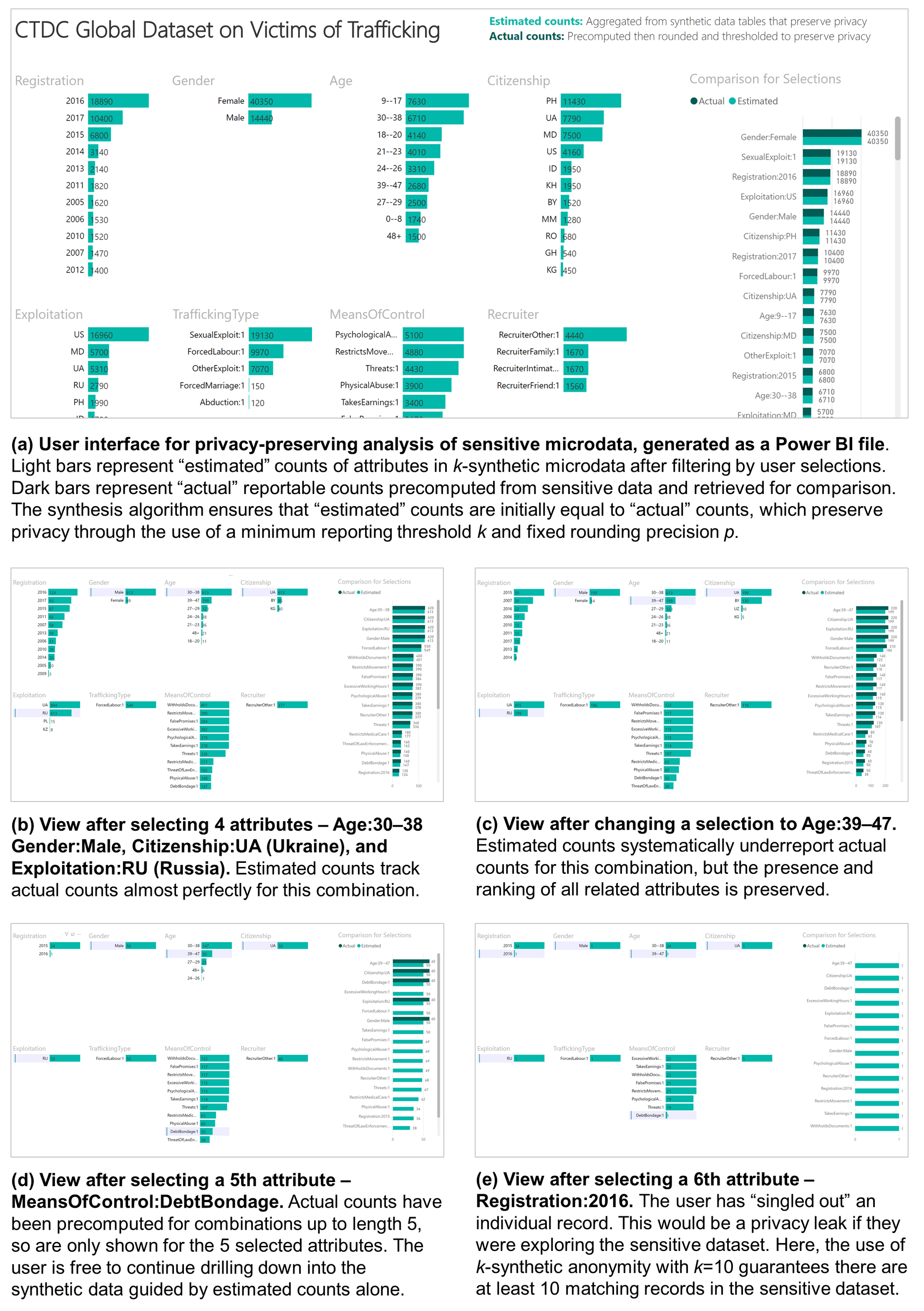}
  \caption{Privacy-preserving analysis of sensitive microdata using synthetic data, aggregate data, and user interfaces generated by our pipeline. Provides a new data platform for the Counter-Trafficking Data Collaborative.
\label{demo}}
\end{figure}

Use of the pipeline by CTDC and others in the counter-trafficking domain will also have significant societal impact, enabling publication of data that could never otherwise be published and empowering all stakeholders to view, explore, and make sense of data for themselves. Figure \ref{demo} provides an overview of the privacy platform enabled by our algorithm, interface, and pipeline.

We structure the paper using the stages of the design study methodology \cite{sedlmair2012design} commonly used in the visualization field. First, we present the literature review that we used to \emph{learn} about privacy concepts and technologies, before describing the accelerator program and launch event used to \emph{winnow} potential directions, \emph{cast} project stakeholders in the broader system of counter-trafficking activity, and \emph{discover} the existing practices by which data on victims of trafficking are collected, integrated, protected, and released. We then describe the \emph{design} and \emph{implement} stages of our process and how they led to a new candidate data platform for CTDC. Finally, we outline current results on two victim datasets and plans to \emph{deploy} the revised platform via the CTDC website, before we \emph{reflect} on the implications, limitations, and future directions of the work.

%% file: related_work.tex
\section{Privacy Concepts and Technologies}

Data protection laws such as the EU General Data Protection Regulation of 2016 (GDPR) \cite{GDPR} offer legal definitions of privacy that can inform the design of privacy-preserving technologies. Article 5(1) states that \textit{``Personal data shall be kept in in a form which permits identification of data subjects for no longer than is necessary ... personal data may be stored for longer periods insofar as the personal data will be processed solely for archiving purposes in the public interest, scientific or historical research purposes or statistical purposes ... subject to implementation of the appropriate technical and organisational measures''}. Pseudonymisation (the replacement of identifiers with pseudorandom strings) is given as an example of such a measure in Article 6(4), yet Recital 26 reiterates that pseudonymised data is still personal data on an \textit{``identifiable natural person''}, and Recital 85 gives \textit{``unauthorized reversal of pseudonymisation''} as an example of a personal data breach that requires reporting to the supervisory authority. The bar is high for data to be considered anonymous and therefore beyond the scope of the GDPR, with Recital 26 stating that \textit{``account should be taken of all the means reasonably likely to be used, such as singling out, either by the controller or by another person to identify the natural person directly or indirectly ... taking into consideration the available technology at the time of the processing and technological developments''}. Doing so requires an understanding of what an attacker could learn from statistical disclosure and how such disclosure can be limited.



\subsection{Syntactic anonymity for microdata release}

Syntactic anonymity methods rely on \emph{safety in numbers} as protection against \emph{singling out} -- the idea that the record for an individual cannot be identified within a crowd of sufficiently similar records. The Datafly system \cite{sweeney1997guaranteeing} was an early attempt to systematically control syntactic anonymity by suppressing, substituting, and generalizing attribute values to reach a minimum count of records in the \emph{equivalence class} of records sharing those values. These ideas were formalized by the definition of $k$-anonymity \cite{sweeney2002k, sweeney2002achieving}, which holds whenever the record for an individual cannot be distinguished within an equivalence class of at least $k$ records sharing the same \emph{quasi-identifiers} -- attributes that describe aspects of the individual and whose combination may lead to their reidentification. Common quasi-identifiers include gender, date of birth, and zip code.

While $k$-anonymity is one of the most widely-used privacy techniques, it remains vulnerable to a range of attacks. \emph{Homogeneity attacks} look for instances where an equivalence class of records sharing the same quasi-identifiers also share the same \emph{sensitive attribute} whose disclosure would cause harm to the individual (e.g., political or sexual orientation). $\ell$-diversity \cite{ldiversity} guards against this threat of \emph{attribute disclosure} by enforcing diversity of sensitive attribute values within each class, while $t$-closeness  \cite{li2007t} protects further by ensuring that the distribution of each sensitive attribute within an equivalence class matches the distribution for the full dataset. Both also guard against \emph{background knowledge attacks} in which an attacker reidentifies an individual's record within an equivalence class because of a known sensitive attribute value. However, no syntactic method can guard against background knowledge attacks where a large number of sensitive attributes are known to an attacker, and designating all such attributes as quasi-identifiers can lead to unacceptably high data suppression  \cite{Aggarwal:2005:KAC:1083592.1083696} in ways that limit the statistical value of the data release.

\subsection{Statistical anonymity for microdata release}

Statistical anonymity methods look beyond the distribution of attribute values in the ``sample'' of the dataset to include prior knowledge about the broader population. $k$-map \cite{Sweeney:2001:CDC:935675} generalizes $k$-anonymity such that each tuple of quasi-identifiers in a microdata release correspond to at least $k$ entries in an external population identification database, thus reducing the threat of \emph{identity disclosure} (i.e., record-level reidentification). Similarly, $\delta$-presence \cite{nergiz2007hiding} measures the more general threat of \emph{membership disclosure} in which harm is caused to an individual by using a public dataset to infer their presence in a private dataset, and presents algorithms for achieving such protection.

\subsection{Differential privacy for statistical queries}

A more general form of protection against membership disclosure is provided through the concept of $\epsilon$-differential privacy \cite{Dwork:2006:DP:2097282.2097284, dwork2006calibrating}, which captures the increased risk to the privacy of an individual from participating in a database. The classical approach to achieving differential privacy is to create a database query mechanism that injects calibrated noise into query outputs to mask the impact of any single record. This has been implemented in many ways, including the PINQ \cite{mcsherry2010privacy} extension to the LINQ query language and the Flex \cite{johnson2017practical} database interface supporting statistical SQL queries.



A benefit of differential privacy query mechanisms is that the privacy loss associated with each query can be quantified mathematically. A challenge is that these losses accumulate with successive queries, and systems must stop answering queries once a predefined privacy budget is reached. How to set, manage, and reset privacy budgets are complex policy questions without any accepted standards. The PSI system \cite{454121} is one system that attempts to help users understand the implication of different privacy parameters and the allocation of privacy budgets across multiple queries and users, but the core challenge of an exhaustible privacy budget remains. Related work \cite{643747} also attempts to address the serious statistical biases that can be introduced through differential privacy mechanisms, which may lead to false inferences detrimental to the subject population and society.

\subsection{Synthetic microdata release}

An alternative approach to microdata release is to synthesize a new dataset in which the records do not correspond to actual individuals, but which preserve the structure and statistics of the original data. Rubin first proposed the concept of synthetic microdata \cite{rubin1993statistical} as an extension of the multiple imputation method used to fill in missing data values based on conditional probability distributions. They highlighted the guarantees that could be made to data subjects that their data would never be shared directly, as well as the guarantees to data analysts about the fidelity of the synthetic data, the ability to use standard tools for its analysis, and the potential to submit analyses prepared on synthetic data for private evaluation by the controllers of the sensitive data.

Modern machine learning methods be easily be adapted to such generation of synthetic data. For example, cross-sampling with decision tree and naive Bayes classifiers has been used to generate synthetic records whose quasi-identifiers are preserved from sensitive ``seed'' records and whose sensitive attributes have no dependence on those of their seed record \cite{LIU2019421}. 


\subsection{Differential privacy for synthetic data release}

Perturbation of microdata has also been shown to achieve differential privacy if the perturbation mechanism can be represented as misclassification matrix that contains no zeros \cite{sholmo2012privacy}. Differential privacy mechanisms can also be used to produce fully synthetic data for release, including contingency tables and OLAP cubes \cite{barak2007privacy}. This approach adds Laplace noise to the Fourier projection of the source table before projecting back to create a synthetic table in the integer domain. Any subsequent queries or operations on the synthetic table do not access the raw data and thus do not cause additional privacy loss. PriView \cite{qardaji2014priview} uses the alternative approach of maximum entropy optimization to support $k$-way marginal contingency tables for high-dimensional datasets.

Several methods have also been proposed that construct a differentially-private model of sensitive data and then sample from that model to construct synthetic microdata for release. DPSynthesizer \cite{li2014dpsynthesizer} uses differentially private one-dimensional marginal distributions and gaussian copula functions to model attribute distributions and their interdependence. PrivBAYES \cite{zhang2017privbayes} works similarly but with low-dimensional marginal distributions and Bayesian networks respectively. This approach allocates the privacy budget to learning pairwise correlations between attributes, but this does not scale to high-dimensional data. Other work \cite{chen2015differentially} proposed a sampling and thresholding mechanism for learning such pairwise correlations without dividing the privacy budget in proportion to $n \choose 2$. Under looser constraints, DPPro \cite{xu2017dppro} uses random projections that maintain probabilistic $(\epsilon,\delta)$-differential privacy \cite{dwork2006our}. The same privacy guarantees have also been recently demonstrated for synthetic data produced using deep generative models in the form of both auto-encoders \cite{abay2018privacy} and generative adversarial networks \cite{frigerio2019differentially}, extending privacy-preservation to multimedia data. 

While differential privacy by definition masks the presence of any individual in a dataset, this protection does not extend to groups of individuals and may still allow membership inference based on the numbers of individuals described by the same combinations of attributes. The concept of \emph{plausible deniability} \cite{bindschaedler2017plausible} adds such group-level protection to probabilistic $(\epsilon,\delta)$-differential privacy by requiring that any output record could have generated from any of $k$ seed records with similar probability by which it was generated from its own seed. However, when the sensitive data are high-dimensional and sparse, either the level of plausible deniability or data utility must decrease because of the randomization necessary to maintain such deniability. 


\subsection{Privacy-preserving visual analytics}

In addition to the many different approaches to data anonymization, there is also a small body of published work on privacy-preserving visualization and visual analytics. Visual representations can themselves preserve privacy based on the inherent ambiguity of spatial aggregations, for example in privacy-preserving parallel coordinates \cite{dasgupta2011adaptive}, scatterplots \cite{dasguptaguess}, and sankey diagrams \cite{chou2019privacy} that apply $k$-anonymity \cite{sweeney2002achieving, sweeney2002k} and $\ell$-diversity \cite{ldiversity} to the geometry of clustered data points. Related work presents a range of privacy and utility metrics for the evaluation of such cluster-based representations \cite{dasgupta2013measuring}. A variety of approaches have also been explored to create privacy-preserving heatmaps of location trajectories, including privacy-preserving user count calculation and kernel density estimation with and without a user diversity index \cite{oksanen2015methods}.

A limitation of applying privacy-preserving methods at the visualization layer is that such methods typically require access to the full sensitive dataset. From a collaboration perspective, this is problematic because some sensitive data may never be sharable without prior anonymization, and any data shared without such protection remains vulnerable to security breaches and privacy leaks. An alternative approach is therefore to create interfaces that allow users to visually explore the trade-off between privacy and utility resulting from different combinations of anonymization methods -- an idea that has been applied to both tabular data \cite{wang2017utility} and graph data \cite{wang2018graphprotector}.










%% file: discover.tex
\section{Analysis of Data Sharing Challenges}

The accelerator program began in July 2019 with a launch event structured as a two-day workshop. Participants included CTDC, experts from diverse backgrounds including law enforcement agencies, counter-trafficking organizations, survivors of trafficking, Tech Against Trafficking (TAT) member companies, and TAT's research, advisory, and support network. This group includes a range of international organizations including Business for Social Responsibility (BSR), the RESPECT initiative, the Global Initiative Against Transnational Organized Crime, GSMA, IOM (UN Migration), the Organization for Security and Co-operation in Europe (OSCE), techUK, University College London, UNSEEN, and the World Business Council for Sustainable Development. Over the course of the two days, participants formed teams and developed action plans to tackle the key problems faced by CTDC, with safe sharing of victim case records the most urgent need.

During the launch event, an activity-centered design process \cite{8019880} was used to structure and guide team discussions about the activity system that our work aimed to transform. This process, grounded in Engestr{\"o}m's system-oriented approach to Activity Theory \cite{engestrom1987learning}, organizes concepts from the target activity and identifies the tensions that characterize the structure and dynamics of that activity. As a seed for our discussion, we began by analyzing two problem statements prepared by CTDC before the event (highlighting key concepts in italics):

\textbf{How can CTDC data on identified victims of trafficking be used to combat trafficking?} CTDC’s mission is to develop the availability of \emph{data and evidence} for \emph{counter trafficking programs} and to provide a mechanism for organizations to move data to \emph{public and policy audiences}. Through the accelerator, CTDC seeks to further develop its \emph{partnership process} and explore and understand the ways in which data on \emph{identified victims of trafficking} can be used to \emph{combat human trafficking}.

\textbf{How can CTDC data on identified victims of trafficking be shared effectively with concerned stakeholders?} CTDC’s current \emph{ad-hoc solutions} are often \emph{labor intensive} or \emph{partner reliant} and there may be scope for improvement. Because of the \emph{sensitivity of the data} published, one key area of concern is \emph{anonymization}. If publicly available data is not correctly anonymized, others may be able to identify those involved. CTDC currently ensures that data is anonymized through \emph{k-anonymization}. However, the process to do so results in the loss of much potentially useful and crucial data. Therefore, CTDC is currently exploring other options to \emph{share more data} and allow \emph{more effective research} to be done while still \emph{protecting privacy and civil liberties}. However, CTDC does not have expertise in implementing \emph{differential privacy} and is worried about the \emph{costs}.

\subsection{Products}

Products are the different types of outcome that motivate the activity. In our case, the multiple products of activity were interrelated: \emph{combating human trafficking} by \emph{supporting effective research} by \emph{publishing victim of trafficking data} while \emph{protecting privacy and civil liberties}.

The tension from the literature of \emph{data privacy vs. analytic utility} captures the implied challenge of supporting effective research to combat trafficking (high utility bar) by publishing data on victims of trafficking (high privacy bar). This overarching tension is reflected all across the activity system, and in general the goal is to develop techniques that achieve high levels of both privacy and utility. Analytic utility can also be thought of as partly inherent in the preservation of sensitive data qualities in the data release (e.g., structure and statistics), and partly contingent on the context in which that data release is accessed and analyzed.  Data visualization and visual analytics both have a significant role to play in making trafficking data accessible and usable by all stakeholders.  

\subsection{Personas}

Personas are the different types of people using the tools of the activity. CTDC participants in the accelerator program and launch event represented the \emph{front-line analyst} and \emph{gatekeeper} personas respectively that play a critical role in design study methodology \cite{sedlmair2012design}. While the front-line analyst was responsible for all forms of data preparation and publication -- spanning microdata anonymization, dashboard construction, and data story production -- the gatekeeper was the primary data custodian responsible for technical oversight and project management as well as partnerships and stakeholder engagement. The key tension here was \emph{ease of application vs. ease of justification} for the privacy mechanisms applied to case records and their impact on analytic utility. Visual tools that can be evaluated interactively demonstrate ease of application and are more easily justified than non-visual tools (e.g., algorithms presented independently of user experience).

\subsection{Capabilities}

Capabilities represent tool support for different types of task. The problem statements highlighted \emph{anonymization} as a crucial task for \emph{developing the availability of data and evidence}. The view of CTDC participants was that the current anonymization mechanism for release of the CTDC global dataset ($k$-anonymization with $k=11$ over the quasi-identifiers of age, gender, and citizenship) resulted in a large loss of utility from data suppression. This took the form of both algorithmic suppression by the $k$-anonymization process, which removed 40\% of the total records, and elective suppression of many valuable data columns that were conservatively judged as having potential for reidentification when used in combinations that could not be fully anticipated. 

A walk-through of the CTDC website also revealed \emph{visualization} to be an important channel for sharing evidence in the form of interactive dashboards and data stories, created using a combination of embedded interfaces developed in Microsoft Power BI, Google Maps, ArcGIS, and DKAN. While these visualizations were built on top of the full database of deidentified case records to create accurate reportable statistics, they were \emph{labor-intensive} to produce because of the \emph{ad-hoc} way in which analysts had to manually filter out rare (and thus potentially linkable) attribute combinations. Because successive ``drill-down'' selections can rapidly filter data down to very small subsets, these dashboards were often constructed to allow filtering on just a single attribute rather than allowing open-ended exploration. This negative impact on analytic utility was also accompanied by an inconsistency in the statistics derived from the $k$-anonymized CTDC global dataset download and online reports (dashboards and data stories) based on the full victim database. 

Overall, these challenges reflect a tension between \emph{releasing datasets vs. releasing data reports}. Both are necessary for different users and use cases, and an ideal release mechanism would combine both in a consistent way accessible through the visualization tools already in use. 

\subsection{Contexts}

Contexts are the different types of external factor that shape the activity. The most salient factor in the target activity system is that a single member organization is the data custodian for CTDC, responsible for integrating and publishing data on behalf of the collaborative. This organization is thus \emph{partner reliant}, dependent on the capacity of other data providers (e.g., NGOs working directly with trafficking survivors) to make regular contributions to the global dataset. Limited capacity to engage in legal data sharing agreements with \emph{counter-trafficking programs} and other potential data users places the onus on this organization to collect, integrate, anonymize, and publish updated data on a regular basis, with sufficient utility to support correct data inferences and effective real-world interventions. We summarize this tension as a \emph{provider driven vs. user driven} release cadence. While superior privacy could attract new data providers, superior utility could similarly attract new users and use cases. Note that in both instances it is not enough to be technically superior -- the anonymization mechanism, the privacy guarantees it provides, and the utility of the anonymized data must all be communicated in clear and understandable terms.  

\subsection{Roles}

Roles are the different types of coordinated contribution to the activity. Users of published data, dashboards, and evidence play a significant role in the overall activity system. The problem statements called out \emph{public and policy audiences}, with surveys on the CTDC website indicating that the main audience is academic researchers (62\%), followed by NGOs (11\%), public sector practitioners (7\%), and international organizations (7\%). At the launch event, representatives from law enforcement and business supply chain management, as well as survivors of trafficking, all advocated for their roles as stakeholders in counter-trafficking data collaboration. A tension in supporting the needs of all stakeholders is their differing \emph{case orientation vs. problem orientation}. Data providers typically work directly with victims and the natural data format for them is the individual case record. Such microdata is also the natural input format for visualization tools used to construct aggregations and distributions for analysis. The majority of data users are more interested in the high-level trends and patterns that result, rather than the precise contents of individual records. 

\subsection{Rules}

Rules are the different types of constraint on the performance of the activity. In the case of publishing data on victims of trafficking, we can reframe the rules that must be followed as the risks that must be mitigated. The need to minimize (if not eliminate) these risks succinctly captures the high-level design requirements for new tools: (1) the \emph{privacy risk} of data subjects being linked to a published record or dataset; (2) the \emph{utility risk} of data users making false inferences and reports from data transformed to reduce privacy risk; and (3) the \emph{accessibility risk} of stakeholders not being able to evaluate how privacy and utility risks are controlled, or analyze data using the tools provided.

These risks also suggest their own tension as a guiding principle for design: the need for \emph{technical guarantees vs. acceptable guarantees}. For example, while techniques based on differential privacy might be able to offer strong mathematical guarantees about the level of privacy loss, in practice such levels might be unacceptably high or lead to unacceptable loss of analytic utility. Guarantees of privacy or utility that are presented in overly technical terms may also be opaque to stakeholders whose informed consent is crucial to the practical and ethical sharing of data. Conversely, techniques like $k$-anonymization may be acceptable despite their weaker technical guarantees because they are easy to understand and apply while meeting legal definitions of deidentified data. In the following section, we present new privacy-preserving mechanisms designed to maximize such acceptability.

%% file: design.tex
\section{Design of a Privacy-Preserving Data Platform}
\label{design}

Our design challenge was to translate the risks identified through our discovery process into appropriate privacy-preserving mechanisms informed by our literature review. Our corresponding design process was highly iterative, exploring new and existing techniques on representative victim of trafficking data (the CTDC global dataset) and evaluating results with key stakeholders. Through this process, \emph{attribute combinations} emerged as a critical concept for risk management:

\begin{itemize}
    \item managing \emph{privacy risks} by controlling the attribute combinations that can appear in the records of a microdata release;
    \item managing \emph{utility risks} by releasing reportable aggregate counts of cases matching different attribute combinations (i.e., queries);
    \item managing \emph{accessibility risks} by enabling interactive visual exploration and evaluation of these complementary datasets.
\end{itemize}

\subsection{Managing privacy risks with $k$-synthetic anonymity}

The principal risk to privacy is traffickers operating according to the prosecutor model \cite{prasser2015putting}, seeking to reidentify specific victims in the published dataset based on distinguishing combinations of attributes. The trafficker must first be able to link a combination of attributes to the victim using background knowledge on their victims and how they were trafficked. Second, they must believe that this combination is rare within the population of all victims. Third, the combination must be unique in the published dataset for the trafficker to confidently link the victim to a specific record.

The risk of identity disclosure can be managed through the use of \emph{synthetic data} in which records no longer correspond to actual individuals. However, if the synthesis mechanism reproduces sensitive attribute combinations and the trafficker can identify combinations that are rare in the dataset, rare in the population, and linkable to known victims, they could infer that the victim is present in the dataset. Such membership inference would also be reasonable, since published data that misrepresents such combinations could be damaging (e.g., suggesting false trafficking routes).

A direct solution is to adopt equivalence class constraints, similar to $k$-anonymity, that control the combinations of attributes which may appear in the records of synthetic data releases. Such constraints can be applied to the results of any data synthesis method, including those offering differential privacy (e.g., \cite{zhang2017privbayes, chen2015differentially, xu2017dppro}). In contrast with the probabilistic guarantees of differential privacy, however, such constraints on counts are concrete, easy to understand, and capable of masking the presence of \emph{groups}, not just individuals. In the context of human trafficking case records, they are also easy to justify in terms of addressing the risk of traffickers inferring the presence of victims in the sensitive dataset. This is not just a privacy risk, but a safety risk -- such beliefs may lead to retaliation against the victim for collaborating with case workers and the implied likelihood of collaboration with law enforcement. Such retaliation may be targeted directly at the victim or indirectly at their close friends and family, and may lead to physical and psychological harm in addition to the original crime.

We combine both of these concepts into the new notion of \emph{k-synthetic anonymity} that generalizes the notion of $k$-anonymity to all columns of a sensitive dataset, with the guarantee that all combinations of attributes appearing in the records of a derived synthetic dataset are frequent ($count \ge k$) in the sensitive dataset. This guarantee preserves the relationships between attributes and prevents the unwanted implication of unobserved or rare (i.e., potentially disclosive) relationships.

\subsection{Managing utility risks with reportable aggregate data}

Regardless of the theoretical utility of synthetic data, if users do not have confidence in the accuracy of statistics derived from synthetic data then they may not be willing to report them. Conversely, if users derive and report inaccurate statistics from published synthetic data, the broader audience of stakeholders within the data sharing ecosystem may lose confidence in the data publisher. The risks of non-reporting or misreporting of victim statistics are also significant in terms of the potential impact on decisions made, resources allocated, and policies implemented to combat trafficking.

Major international reports on human trafficking typically report only high-level statistics. For example, in the 2018 Global Trafficking in Persons Report by the UNODC (United Nations Office on Drugs and Crime) \cite{unodc2018}, statistics included number of detected victims by year and region, share of detected victims by region of origin and detection, shares of detected victims by age group, sex, and region, and forms of exploitation by region. The CTDC website also offers visualizations and data stories showing distributions of case attributes by region, industry, sex, and age group. The implication is that publishing the aggregate counts of cases matching small combinations of attributes alongside any microdata release would support the complementary tasks of (1) discovering high-level statistics for reporting, (2) exploring the low-level structure of case records for insights, and (3) using microdata as input to data science or machine learning that requires such a format.

Although the greatest utility is achieved through the publication of precise aggregate counts, the publication of small counts or small differences in counts between successive releases can both be disclosive. The use of a \emph{minimum reporting threshold} can address the risks associated with small counts, while the use of a \emph{fixed rounding precision} can enforce minimum differences over time.

Our notion of reportable aggregate data describes the publication of aggregate counts for the short combinations of attributes ($1 \le length \le \ell$) typically reported in the literature on trafficking, where these counts have been subjected to a minimum threshold $t$ and rounding precision $p$ to avoid disclosing small or precise counts. While high-utility synthetic data should accurately approximate these counts, the publication of reportable aggregate data alongside synthetic microdata removes any uncertainty (and resulting lack of confidence) associated with the use of synthetic data only.

\subsection{Managing accessibility risks with visual analytics}

The need for privacy-preserving visual analytics interfaces was suggested both by the existing use of visualizations on the CTDC website and our proposed publication of two complementary datasets in need of interactive, user-directed comparison. Mainstream Business Intelligence (BI) platforms like Power BI and Tableau offer the potential for \emph{exploratory data analysis} -- analysis that is not driven by prior beliefs, but by the desire to discover meaningful structure in data \cite{tukey1976exploratory}. Such exploratory data analysis is often facilitated through the use of interfaces that follow the information seeking mantra of ``overview first, zoom and filter, then details on demand'' \cite{shneiderman1996eyes}. 

For synthetic microdata, dashboard interfaces can be constructed that show the distribution of values for each data attribute using ``slicer'' visuals that are mutually filtering. As in Figure \ref{demo}, for example, the interface might show an \emph{overview} that juxtaposes visuals for each attribute of the synthetic microdata, with each visual showing the distribution of that attribute by listing its values from most to least frequent. The user can then \emph{zoom and filter} by selecting attribute values, with the effect of filtering the underlying dataset to include only records containing the selected attributes. Multiple selections construct a compound filter that shows both the distributions of related attributes and the  possible ways to extend the filter combination -- offering an ``information scent'' \cite{pirolli1999information} that guides exploration. However, whereas conventional visual analytics is grounded in real records whose \emph{details on demand} may lead to insights about an individual record, this is not the case for synthetic microdata in which each record represents a ``statistical individual'' rather than an actual person (or an ``identifiable natural person'' as described in the GDPR). 

What is of interest during exploratory analysis of synthetic microdata is how the ``estimated'' counts formed by filtering and aggregating synthetic records compare to the ``actual'' counts that would have been seen had the original sensitive dataset been used. This is where the complementary dataset of reportable aggregates achieves its purpose, with actual counts being juxtaposed with synthetic counts whenever they have been precomputed for the current combination of filtering attributes. By precomputing the remaining counts of all attributes after filtering by attribute combinations of length <$\ell$, $\ell-1$ becomes the limit on how many concurrent selections the user may make while retaining the ability to see estimated and actual counts juxtaposed for comparison. Any selections up to this limit can dynamically retrieve reportable values from the aggregate data, while selections made beyond this limit will allow further exploration of the synthetic microdata only. Unlike the privacy budget for queries under differential privacy, this limit is reusable and does not in itself represent a privacy-preserving mechanism (since the thresholding and rounding of aggregate counts could theoretically protect all lengths of attribute combinations).

In a field that has dedicated most visualization efforts towards the needs of data controllers, e.g., understanding the implications of privacy budget allocation in PSI \cite{454121} and DataSynthesizer \cite{ping2017datasynthesizer, howe2017synthetic}, our approach to parallel exploration of complementary privacy-preserving datasets is distinct in its focus on increasing accessibility for diverse users of datasets, not just their creators.

%% file: implementation.tex
\section{Implementation of a Turnkey Pipeline}

In this section, we describe our implementation of an integrated pipeline that transforms a file of sensitive microdata (in CSV or TSV format) into the synthetic microdata and reportable aggregates necessary to drive a generic interface template built as a Microsoft Power BI \cite{powerbi} report. There are many advantages to developing privacy-preserving interfaces within an established visual analytics tool, including familiarity, flexibility, and reliability. However, such tools typically assume the availability of individual-level microdata, which is precisely what cannot be shared (in this case and many others) for privacy reasons. Our implementation overcomes this challenge by supporting visual exploration of synthetic data ``corrected'' in real-time by precomputed aggregate data.

\subsection{Configuring the pipeline}

Our pipeline is written in Python and configured using parameters controlling the generation of synthetic microdata (for minimum reporting threshold $k$) and reportable aggregate data (for minimum reporting threshold $k$, fixed reporting precision $p$, and maximum combination length $\ell$). 

Input data takes the wide format of one row (record) per individual, with multiple categorical attribute columns per row. Single columns are used to represent single-valued attributes (e.g., year of registration), while multi-valued attributes (e.g., means of control) are represented by multiple binary columns.  Continuous numeric attributes (e.g., age) must first be quantized into discrete categories (e.g., age bands) based on the desired level of granularity for reporting and the desire to maintain above-threshold counts for each category/combination.

By default, our approach to controlling the release of attribute combinations applies only to values indicating the presence of an attribute, i.e., not zero or null values indicating its absence. An additional pipeline parameter allows the listing of any columns where zero values may potentially be identifying -- such values are then included when creating and counting attribute combinations.

\subsection{Generating synthetic microdata}

Our algorithm for generating data with $k$-synthetic anonymity is designed for the specific use case of exploring counts of records matching combinations of selected attributes. It embodies the idea of \emph{attribute conservation} to ensure utility at the level of individual attribute counts, which we achieve by preserving all attribute values present in the sensitive data but manipulating their distribution across records to ensure $k$-synthetic anonymity. Since preserving precise attribute counts is itself a privacy risk, however, we take the same notion of a reporting precision used to create aggregate data and use it to manipulate the implicit counts of attributes in synthetic microdata such that these imprecise counts match precisely. The high-level stages of our synthesis algorithm are as follows:

\begin{enumerate}
    \item \emph{Sample synthetic records from sensitive seeds}. Build synthetic records by progressively sampling attributes (without replacement) from corresponding ``seed'' records in the sensitive dataset, stopping in each case when further sampling would break $k$-synthetic anonymity. Maintain counts of all remaining attribute values that were not sampled from seed records.
    
    \item \emph{Update remaining counts based on reporting precision}. Add or subtract counts of attributes remaining such that attribute totals (across the synthetic records and counts remaining) match the aggregate values reported after thresholding and rounding.

    \item \emph{Sample synthetic records from remaining attributes}. Build synthetic records by progressively sampling attributes (without replacement) from the counts of attributes remaining, starting a new record whenever further sampling would break $k$-synthetic anonymity.

    \item \emph{Suppress excess attributes in synthetic records}. Deal with negative counts of attributes remaining by randomly suppressing these attributes from all synthetic records created.
    
    \item \emph{Sort and save synthetic records}. Sort synthetic records first by their natural order and second by their length, preventing leakage of information about their corresponding order in the sensitive dataset and enabling simple visual examination of record groups. Save as output.
    
    \item \emph{Report the synthesis ratio}. Calculate and report the \emph{synthesis ratio} as the number of synthetic records divided by the number of sensitive records. This ratio communicates the difficulty of achieving $k$-synthetic anonymity given the nature of the data and the given value of $k$ (higher values indicate more record breaks which reflect greater anonymization difficulty).

\end{enumerate}

The generation of synthetic microdata without the use of seed records is also supported by our pipeline. In this case, synthetic records are instead created by sampling from conditional attribute distributions until further sampling would break $k$-synthetic anonymity. While such unseeded synthesis may create longer records on average and better preserve structure for machine learning, seeded synthesis is generally faster and better preserves statistics for visual analytics.

\subsection{Generating reportable aggregate data}

Our pipeline precomputes the counts of all attribute combinations with lengths $\le \ell$, with the precise actual counts protected through the use of a fixed reporting precision $p$ and the same minimum reporting threshold $k$ used for data synthesis (applied both before and after rounding to the closest multiple of $p$). These privacy-preserving counts of attribute combinations are released as a TSV file.

\subsection{Generating visual analytics interfaces}

To accommodate the generation of visual analytics interfaces for any valid input data, our pipeline needs to allow for variability in the number of input columns, the mappings from input columns to output visuals, and the groupings of visuals into pages used to answer different analytic questions. To support this flexibility, we developed a generic, single-page interface template within a Power BI Desktop report (Figure \ref{demo}) that can be manipulated programmatically based on input data and configuration parameters. The template page comprises a page title, a grid of Attribute Slicer \cite{attributeslicer} visuals prepared to rank attribute values by the count of corresponding records, and a combined list of all attribute values comparing ``Estimated'' counts (from dynamic aggregation of the synthetic microdata table) against ``Actual'' counts (from dynamic lookups into the reportable aggregates table). Estimated and Actual are both implemented as Data Analysis Expressions (DAX) measures \cite{dax} generated by our pipeline to match the columns of the synthetic microdata to be visualized.

The user can specify a list of titled pages comprising lists of visuals, with each visual bound to either a single column or a named list of related columns (e.g., all binary columns expressing a multi-valued attribute).  Our pipeline programmatically populates a grid of Attribute Slicer visuals based on the visuals specified for each page. The resulting interface is ready to use and may be freely shared as a privacy-preserving interface to inherently anonymous data, either directly as a Power BI Desktop file or via publication to the Power BI service for organizational or public access.

\subsection{Evaluating utility and privacy}

Utility is measured in terms of how well synthetic attribute counts preserve the value of their corresponding sensitive attribute counts, given that the breaking of sensitive records into multiple synthetic records necessarily ``loses'' the disclosive attribute combinations spanning break points. Synthesizing records from remaining attributes is rarely sufficient to exceed original combination counts. Privacy is guaranteed by design and can be confirmed empirically by observing zero leakage of sensitive attribute combinations that are rare in the sensitive dataset. The pipeline publishes summary TSV data files and SVG data graphics describing both datasets: 

\begin{enumerate}
    \item \emph{sensitive\_rare\_by\_length} -- the count of combinations and rare combinations for each combination length in the sensitive dataset;
    \item \emph{synthetic\_leakage\_by\_length} -- the count of combinations and leaked rare sensitive combinations for each combination length in the synthetic dataset;
    \item \emph{synthetic\_preservation\_by\_length} -- the mean count of filtered records and mean proportion of sensitive count preserved for each combination length in the synthetic dataset;
    \item \emph{synthetic\_preservation\_by\_count} -- the mean length of attribute combination and mean proportion of sensitive count preserved for bins of combination counts in the synthetic dataset.
\end{enumerate}

%% file: discussion.tex
\section{Application to Human Trafficking Datasets}

We now analyze example runs of our pipeline on two datasets representing victims of  trafficking.

\subsection{Public dataset on trafficking victims}

The first dataset is a version of the CTDC global dataset already $k$-anonymized with $k=11$ over the quasi-identifiers of age, citizenship, and gender. This dataset has 55k rows and 33 columns (1.8M cells, 20\% non-zero). Other attributes represent the country of exploitation, the year of registration, and the multiple trafficking types (e.g., sexual exploitation, forced labor, forced marriage), recruiter relationships (e.g., family, friend, partner), and means of control (e.g., debt bondage, threats, movement restriction) that may be associated with each case. The relative sparsity of this dataset results from the majority of columns encoding the presence or absence of an attribute (e.g., a type of trafficking or means of control) in binary form. 

\begin{figure}[h]
  \centering
  \includegraphics[width=.9\columnwidth]{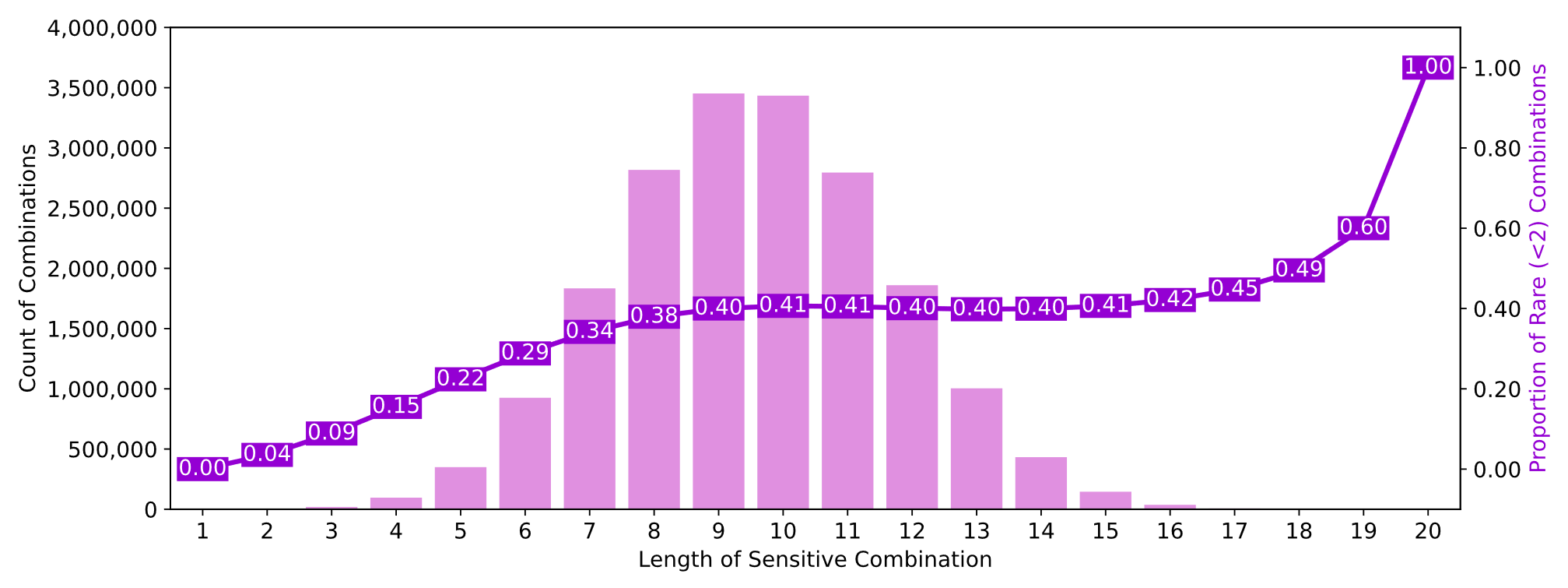}
  \caption{Many unique combinations remain in the sensitive dataset despite prior $k$-anonymization.\label{ctdc_unique}}
\end{figure}

Analyzing this sensitive dataset revealed the presence of many unique (and thus potentially identifying) records despite the prior use of $k$-anonymization (Figure \ref{ctdc_unique}). The proportion of rare combinations increases dramatically as the definition of rare (<$k$, the reporting threshold) increases. For $k=10$, the majority of attribute combinations are rare for each combination length greater than four (reaching 60\% for length 5, 80\% for length 10, and 100\% for length 20).

\begin{figure}[h]
  \centering
  \includegraphics[width=.9\columnwidth]{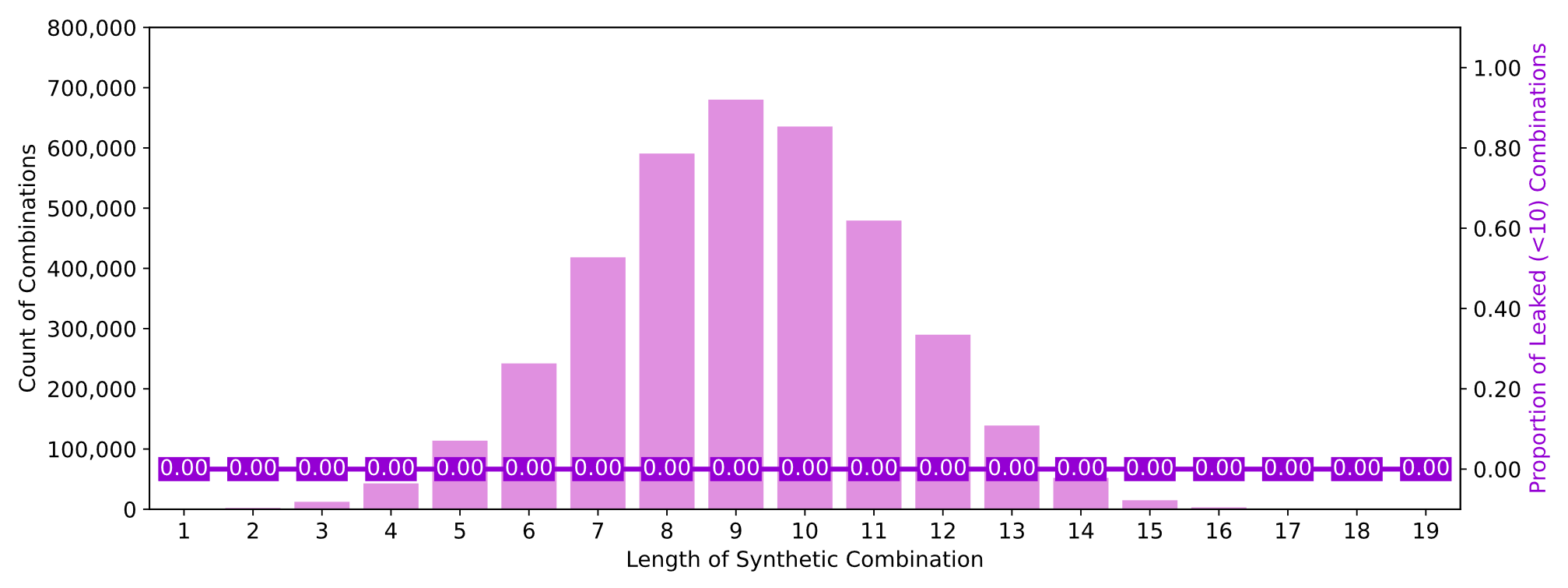}
  \caption{No combinations that are rare (<10) in the sensitive dataset appear in the synthetic dataset.\label{ctdc_leakage}}
\end{figure}

Applying our data synthesis pipeline with $k=10$ and $p=10$ maintains the relative distribution of combination lengths for substantially fewer combinations overall, and achieves the design goal of preventing leakage of combinations that are rare (<$10$) in the sensitive dataset (Figure \ref{ctdc_leakage}). This is a natural consequence of the shorter records that result from sampling subsets of sensitive records. The synthesis ratio in this example was 1.04, indicating that privacy and individual-attribute utility was achieved through a very small (4\%) increase in the total number of records.

\begin{figure}[h]
  \centering
  \includegraphics[width=.9\columnwidth]{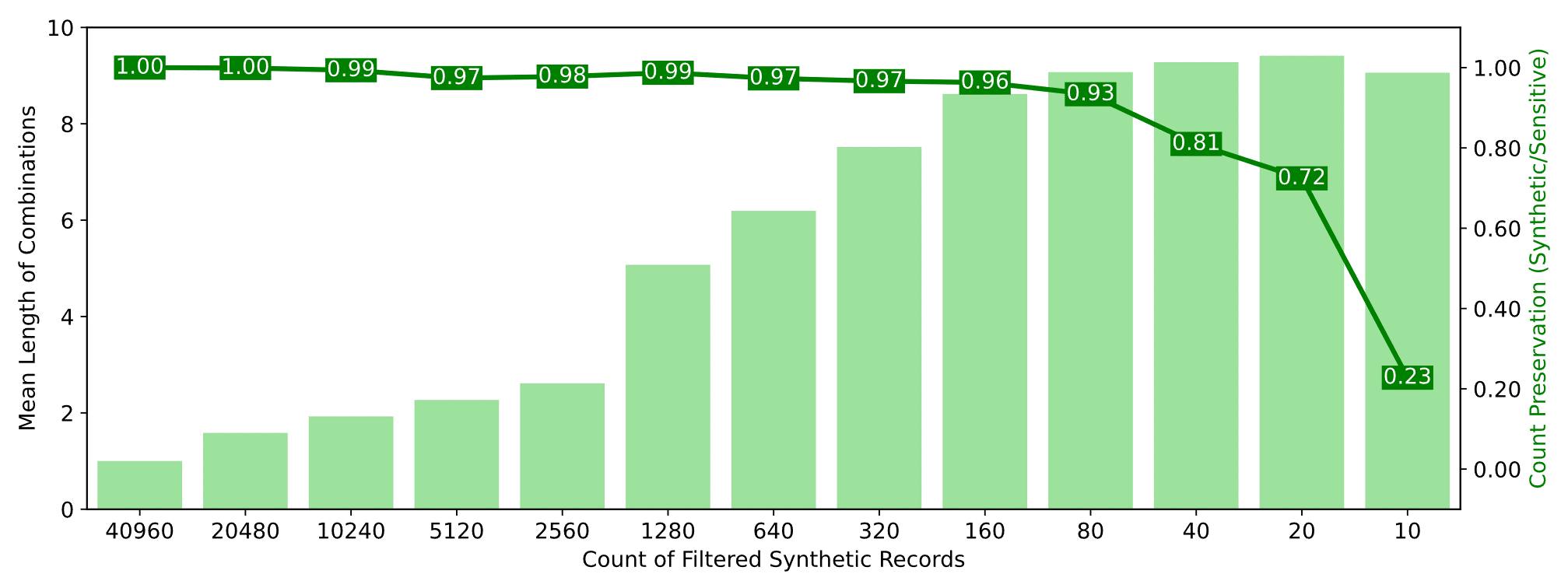}
  \caption{Sensitive combination counts are preserved (>$80\%$) for synthetic combination counts >$20$.\label{ctdc_count}}
\end{figure}

Examining the preservation of sensitive combination counts for synthetic combination counts binned on a logarithmic scale (Figure \ref{ctdc_count}) shows that sensitive combination counts are preserved at a high level (>$80\%$) for synthetic combination counts >$20$ (i.e., down to very small subsets of records). This synthetic data therefore has significant utility for the task of record count estimation even in the absence of comparative aggregate data.  

\subsection{Private dataset on trafficking victims}

\begin{figure}[h]
  \centering
  \includegraphics[width=.9\columnwidth]{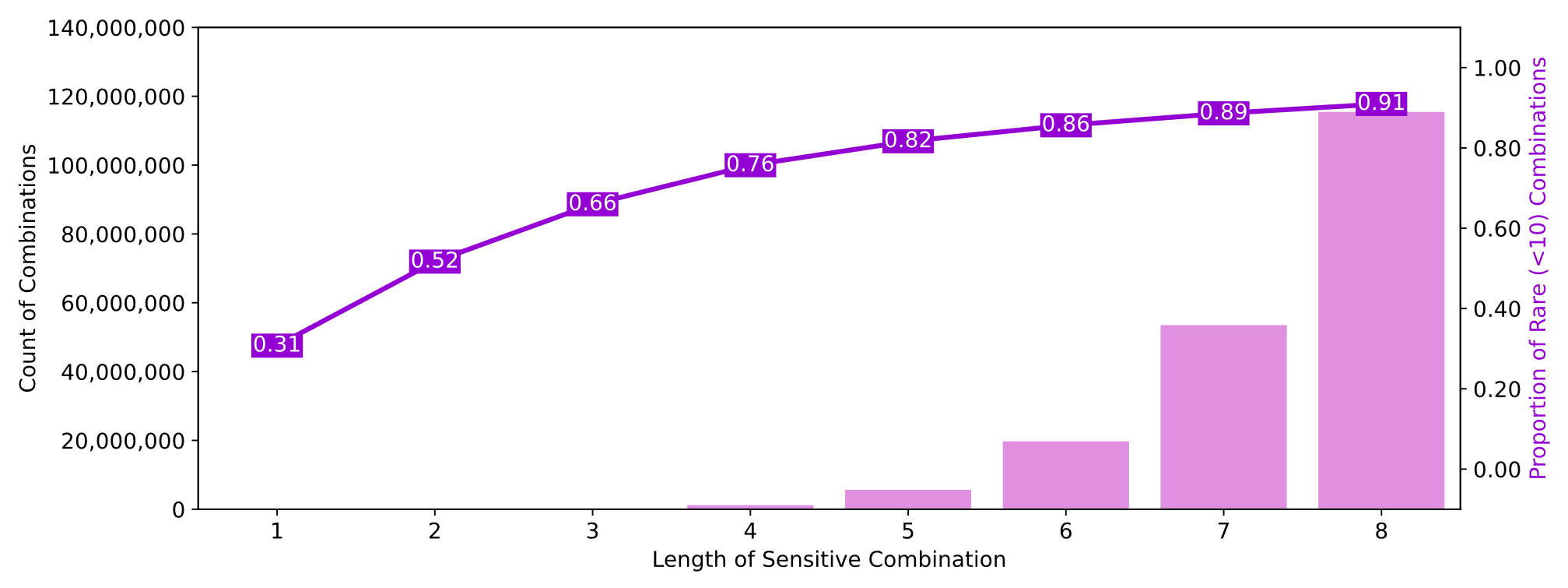}
  \caption{The majority of sensitive attribute combinations are rare for lengths >$1$ (computed with $\ell=8$).\label{iom_rare}}
\end{figure}

The second dataset is an unpublished and deidentified (but not $k$-anonymized) contribution to the CTDC global dataset shared by a CTDC member organization under legal agreement. This dataset has 52k rows and 41 columns (2.1M cells, 22\% non-zero). This dataset adds attributes including the victim's education, marital status, trafficking duration, and children. Despite the apparent similarity to the $k$-anonymized dataset, this dataset exhibits a significantly greater number and proportion of rare attribute combinations (Figure \ref{iom_rare}), representing a major challenge for anonymization.

\begin{figure}[h]
  \centering
  \includegraphics[width=.9\columnwidth]{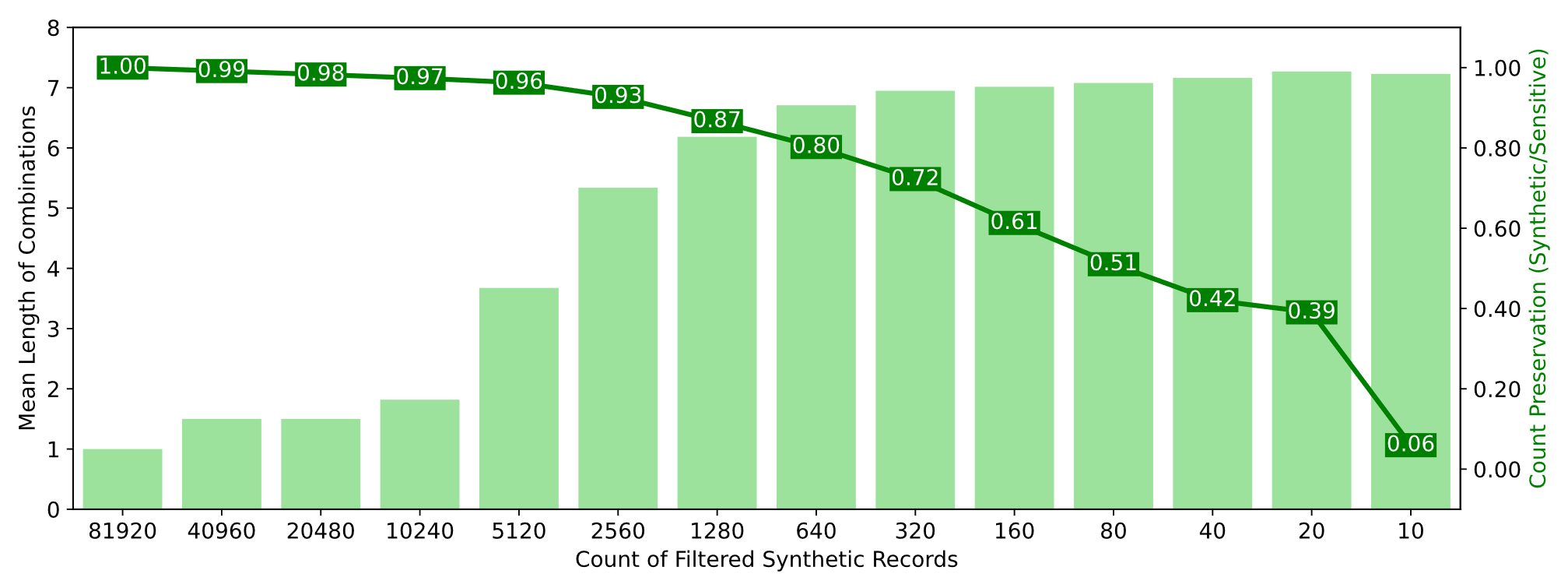}
  \caption{Sensitive combination counts are preserved (>$80\%$) for synthetic combination counts >$320$ ($\ell=8$).\label{iom_count}}
\end{figure}

Applying our data synthesis pipeline with $k=10$ and $p=10$ maintains the relative distribution of combination lengths and achieve zero leakage as with the public dataset. The synthesis ratio in this example was 1.46, indicating that privacy and individual-attribute utility was achieved through a moderate (46\%) increase in the total number of records. Compared with public dataset synthesis ratio of 1.04, this confirms a relatively greater anonymization challenge (as expected from Figure \ref{iom_rare}). The effect on utility, however, is relatively small -- the vast majority (>$80\%$) of sensitive counts are preserved for synthetic counts >$320$, or $6\%$ of the sensitive records potentially of interest (Figure \ref{iom_count}).

%% file: conclusion.tex
\section{Conclusion}

The target beneficiaries of this work are the millions of trafficking victims around the world, the survivors of trafficking, and all the people who might avoid being trafficked through data-driven collaboration within the counter-trafficking community. The solution presented in this paper achieves privacy for all such data subjects, as well as utility for data users. We hope that this rare combination of both privacy and utility will empower data owners to participate in data sharing and collaboration on a scale not previously possible, to help solve societal problems -- human trafficking and otherwise -- that could not otherwise be solved.

Despite the challenges of transforming an existing data platform and working in a complex and sensitive domain, the opportunities are significant. Our notion of $k$-synthetic anonymity automatically prevents the publication of records whose attribute combinations may be used to infer the presence of victims, allowing many more attributes of victim case records to be shared for analysis. Such extra detail could be crucial to understanding aspects of trafficking that are not currently shared due to the potential for accidental privacy leaks. Our proposed publication of reportable aggregates alongside synthetic microdata also aims to ensure that no approach to microdata release, with $k$-synthetic anonymity or otherwise, can override the need for accurate reporting of statistics on which so many decisions, policies, and human lives depend.

In the counter-trafficking domain, CTDC are already using our open-source software to transform the sharing and analysis of victim case records. Reaching this stage has requiring many rounds of iteration in terms of research, design, and development, with each iteration informed by critical review and feedback from diverse stakeholders at CTDC and Tech Against Trafficking. The new privacy-preserving data platform has not yet been released, however, and only with motivated use by the broader counter-trafficking community can we understand the extent to which our solution meets its accessibility goals. The use of interactive visualization and visual analytics as presented in this paper is our first step towards enabling all data stakeholders to view, explore, and make sense of data for themselves -- a critical enabler of informed representation and evidence-based practice.

One of the major design challenges we faced was how users could explore two complementary datasets -- synthetic microdata plus aggregate data -- in parallel. In our solution, users are able to interact with synthetic data as if it were actual sensitive data, with simple juxtaposition of actual and estimated counts whenever such actual counts have been precomputed. This continuous comparison helps users to evaluate the quality of the synthetic data in specific areas of interest that may vary between users (e.g., specific countries or regions), establishing at least some confidence in the accurate co-occurrences, rankings, and counts of attributes when actual counts are not available. Future work may explore (a) how actual aggregates may be used to ``correct'' the representation of synthetic data within visuals themselves, (b) how such visuals may gracefully degrade to synthetic counts and confidence intervals when actual counts are not available, (c) how to automate the selection of pipeline parameters for a given dataset, and (d) how to extend $k$-synthetic anonymity from categorical microdata with one record per individual, to different types of attributes (e.g., continuous numeric and unstructured text attributes) and datasets (e.g., log and graph datasets with one and two individuals per record, respectively). We look forward to exploring these directions with the community via the open-source software repository for this work, accessible via GitHub at \url{https://github.com/microsoft/synthetic-data-showcase}.